\documentclass[preprint, pra, nofootinbib]{revtex4}

\usepackage{epsfig}
\usepackage{amssymb}
\usepackage{epstopdf}
\usepackage{bm}

\begin{document}

\title{Extending Hamilton's principle to quantize classical fields}
\author{K.B. Wharton}
\email{wharton@science.sjsu.edu}
\affiliation{Department of Physics and Astronomy, San Jos\'{e} State University, San Jos\'{e}, CA 95192-0106}

\begin{abstract}
Hamilton's principle does not formally apply to systems whose boundary conditions lie outside configuration space, but extensions are possible using certain ``natural'' boundary conditions that allow action extremization.  With the single conjecture that only such action-extremizing boundaries can be physically realized, the classical relativistic scalar field becomes subject to certain quantization conditions upon measurement.  These conditions appear to be analogous to Bohr-Sommerfeld quantization, and are derived explicitly for the case of angular momentum measurements of a classical scalar field.

\end{abstract}


\maketitle

\section{Introduction}

Equations representing physical events cannot be solved without boundary conditions (BCs), so determining exactly which BCs correspond to which measurements/interactions is of crucial import to all of physical science.  Subtleties concerning BCs generally receive less scrutiny than the equations they constrain, despite well-known open problems.  For example, a spatially extended system requires an initial BC that is also extended, imposed on some particular hypersurface in spacetime.  The problem is that there is no well-defined procedure to determine which hypersurface should correspond to a given physical measurement. 
	 
This Letter explores the consequences of extending Hamilton's principle of extremized classical action to constrain not only the equations of motion, but also the hypersurfaces on which BCs may be imposed.  Classical systems have certain BCs for which no action extremization is possible; a natural extension of Hamilton's principle would imply that such BCs are not physically realizable.  Using only the so-called  ``natural boundary conditions'' that permit action extremization defines a subset of possible hypersurfaces on which BCs can be imposed.  Remarkably, this simple condition leads directly to a quantization of some measurement results on classical fields.

\section{Extending Hamilton's Principle}

Consider a typical formulation of Hamilton's principle: ``\textit{The [action] becomes stationary for arbitrary possible variations of the configuration of the system, provided the initial and final configurations of the system are prescribed.}''\cite{Lanczos}  The final clause restricts Hamilton's principle to problems where initial and final BCs constrain coordinates in configuration space.  For other types of BCs, it is known that the action cannot generally be extremized \cite{Lanczos, CnH, Elsgolc}.  One approach to this problem is to add boundary terms to the action, with different terms for different types of BCs.  While this procedure is mathematically useful, altering the action to permit ``action extremization'' reduces Hamilton's principle to a near-tautology, and diminishes the fundamental role of the Lagrangian.  This paper takes the position that the word ``action'' should always refer to the usual spacetime integral of the Lagrangian density, and solves the problem by restricting the BCs to those that extremize the action (as is also done in the literature \cite{Hebecker, Csaki, Dolce}).

A straightforward extension of Hamilton's principle would simply remove the final clause in the above formulation and require the action to be stationary for \textit{all} BCs, even ``non-coordinate BCs'' (NCBCs) that do not specify points in configuration space.  It is important to note that NCBCs are common and physically meaningful.  For example, consider a precise velocity measurement (say, via a collision) that leaves the position unspecified.  This measurement, together with its result, is a constraint on $\dot{q}$ -- but not the coordinate(s) $q$ -- of a Lagrangian $L(t,q,\dot{q})$.  Such an NCBC-constrained action can only be extremized if the NCBC is also a ``natural'' boundary condition as defined in \cite{Lanczos, CnH}.

For a demonstration of this central point, consider Hamilton's principle as applied to a particle of mass $m$ in one dimension, subject to an unchanging potential $V(x)$.  The boundary conditions are imposed at times $t_0$ and $t_1$.  The variation in the action $S$ can then be constructed from the Lagrangian $m\dot{x}^2/2-V(x)$ as
\begin{equation}
\label{eq:Pds}
\delta S =  \int^{t_1}_{t_0} \left(-V'(x) \delta x - m\ddot{x} \delta x\right) dt + \left.  m\dot{x} \delta x \right|^{t_1}_{t_0}= 0.
\end{equation}
 
The only way to make the integral in (\ref{eq:Pds}) equal zero, for all possible path variations $\delta x$, is if $m\ddot{x}=-V'(x)$; this is the Euler-Lagrange equation (and Newton's second law).  But unless the boundary term in (\ref{eq:Pds}) also vanishes, the solution $x(t)$ that fulfills both the BCs and Newton's second law will \textit{still not extremize the action} under all variations $\delta x$ consistent with the BCs.  The boundary term is indeed zero if the BCs constrain the initial and final values of $x$, because in that case $\delta x=0$ at both $t_0$ and $t_1$.  But if instead $\dot{x}$ is constrained at either boundary, leaving $x$ free to vary, then the boundary term in (\ref{eq:Pds}) is generally non-zero.\footnote{One could imagine imposing additional mathematical BCs on $x(t)$ that would fix the appropriate endpoints, but those BCs would not correspond to external physical constraints.}  For such NCBCs, $\delta S \ne 0$, and Hamilton's principle seems to fail except for the ``natural'' boundary $\dot{x}=0$.

Clearly, non-zero lab-frame velocity measurements are physically realizable.  But it is only measured velocities of exactly zero for which the boundary term in (\ref{eq:Pds}) vanishes and the action becomes stationary.  The previous two sentences can be reconciled with an extended Hamilton's principle by noting that there is always a reference frame in which the measured velocity is zero.  Therefore, there is always a ``natural'' hypersurface which permits action extremization (for any lab-frame velocity) at the expense of the freedom of the hypersurface on which the corresponding BC is imposed.  Given the lack of a well-defined procedure for determining measurement hypersurfaces in the first place, it seems reasonable to consider the following conjecture: \textbf{Any boundary condition inconsistent with an extremized action is not physically realizable}.  This conjecture restricts NCBCs to lie on a class of ``natural'' hypersurfaces, as the above example illustrates. 

Some authors have already implicitly assumed the above conjecture in an extra-dimensional context \cite{Hebecker, Csaki}, and Dolce has recently shown that periodic BCs motivated in this manner may lead to a novel quantization procedure for 4D bosonic fields \cite{Dolce}.  Still, these authors treated the above conjecture as established physics, rather than as a possible new physical principle.  These authors also did not discuss that such a constraint-on-constraints seems inconsistent with the very concept of a ``boundary condition".  In other words, if BCs are external constraints on a system, how could those external constraints be in turn constrained by the system itself?  For a preliminary answer, note that in the above example the natural hypersurface is determined using only BC-constrained parameters (the velocity $\dot{x}$ at the boundary), so there is no dependence on the full trajectory $x(t)$.  This point will be revisited below. 

\section{Implications for classical fields}

It is easy to see that the above conjecture can have no serious implications for the case of a single particle; a localized boundary condition is trivially zero off the particle's world line in \textit{every} reference frame.  Even for multi-particle systems, the complete freedom of the hypersurface between particles allows for so many possible geometries that it is difficult to see how any consequences could result.  But for the case of continuous extended systems, the continuously-restricted BC hypersurface leads to physical consequences.

These consequences are most easily explored using a classical scalar field $\phi(\bm{x},t)$ in Minkowski spacetime, where the Lagrangian density is given by
\begin{equation}
\label{eq:FLag}
{\cal L}=\frac{1}{2}\left[ \frac{1}{c^2}\left(\frac{\partial\phi}{\partial t}\right)^2 -(\nabla \phi)^2 - \frac{m^2c^2}{\hbar^2}\phi^2 \right].
\end{equation}
One can confirm that the corresponding Euler-Lagrange equation is the Klein-Gordon equation (KGE), $\Box \phi + (m^2c^2/\hbar^2)\phi=0$.  Note that although the usual parameters $m$ and $\hbar$ force the KGE dispersion relation to be consistent with the deBroglie relations for a generic particle of mass $m$, the field itself is still a classical field.  ($\phi$ is a real function over all spacetime, and is not operator-valued, so no commutation relationships are imposed.)  Because the sum of any two solutions to the classical KGE is an equally valid solution, there are no quantization conditions on $\phi$.

The covariant generalization of the action requires a boundary condition imposed on a closed hypersurface $\bm{s}$ that surrounds a closed 4-volume $\Omega$.  The below analysis will be restricted to cases where $\bm{s}$ can be split into an initial spacelike surface $\bm{s}_0$ and a final spacelike surface $\bm{s}_1$.  (This reduces to the common two-time case when $\bm{s}_0$ is the flat hypersurface $t=t_0$ and $\bm{s}_1$ is $t=t_1$; in this case $\Omega$ extends to spatial infinity where $\phi \to 0$.)  The variation of the action $S=\int {\cal L} \, d\Omega$ can be calculated in a similar manner as in Section II, yielding
\begin{equation}
\label{eq:FdS}
\delta S=\int_{\Omega} \left[ \frac{ \ddot{\phi}}{c^2} - \nabla^2 \phi + (m^2c^2/\hbar^2)\phi \right]\delta\phi\,d\Omega+\oint_{\bm{s}} \frac{\partial \phi}{\partial \eta} \delta \phi \, d\bm{s},
\end{equation}
where $\partial/\partial\eta$ is differentiation in the direction of the outward normal of $\bm{s}$.  Setting the bracketed term to zero yields the KGE, but this is not sufficient to extremize the action because of the boundary term.  This final term in (\ref{eq:FdS}) is an integral over the closed hypersurface $\bm{s}$. 

Given that the field $\phi$ solves the KGE, the boundary term reveals two ways to satisfy Hamilton's principle; either $\delta \phi=0$ or $\partial\phi /\partial\eta=0$ everywhere on $\bm{s}$.  The former option is always possible in quantum field theory (QFT), because the unitary evolution of the quantum field is effectively a first-order (in time) differential equation.  Any constraint on $\dot{\phi}$ in Lagrangian QFT can be reexpressed as a constraint on $\phi$, forcing $\delta \phi=0$ on $\bm{s}$ for any BC.  But for this classical field the only equation for $\phi$ is second-order in time (the KGE), meaning the quantities $\phi$ and $\dot{\phi}$ are independent on any space-like hypersurface.  If the BC constrains $\dot{\phi}$ (even in combination with $\phi$), then $\delta\phi$ may be non-zero on $\bm{s}$; such a constraint will continue to be called an NCBC.

With an NCBC that corresponds to an initial measurement of the field, the only way to satisfy Hamilton's principle is if the initial space-like hypersurface $\bm{s}_0$ fulfills $\partial\phi /\partial\eta=0$ everywhere on $\bm{s}_0$.  Such a constraint exactly corresponds to the ``natural boundary condition" defined in Courant and Hilbert for general extremization problems.\cite{CnH}  To better visualize these ``natural'' boundaries, consider a reference frame where $\bm{s}_0$ is a flat hypersurface.  Here the ``natural'' condition is simply $\dot{\phi}=0$, which implies a zero 3-momentum density $T_{0i}=\dot{\phi}\,{\partial \phi}/{\partial x^i}$ within the hypersurface.  

For a specific example, consider a measurement of the z-component of the angular momentum.  In classical field theory this quantity is an integral over an instantaneous hypersurface,
\begin{equation}
\label{eq:Lz}
L_z=\int (xT_{02}-yT_{01}) dx dy dz = \int r p_\theta d^3r,
\end{equation}
where $p_\theta=(\dot{\phi}/r)\partial\phi/\partial\theta$ in cylindrical coordinates.  But if this flat hypersurface is a ``natural'' BC, $p_\theta=0$, and the only allowable value of $L_z$ is zero.  

This conclusion is clearly too restrictive, because it used the assumption of a flat hypersurface.  For an example with a non-zero $L_z$, consider a field where both $p_\theta$ and the energy density $T_{00}$ are non-zero constants everywhere.  Here the natural hypersurface must be tilted (into the time direction) with a slope $p_\theta/(cT_{00})$, as shown in figure 1.  Note the inherent discontinuity in the corkscrew-shaped hypersurface.

\begin{figure}[htb]
\centerline{\includegraphics[width=.7\textwidth]{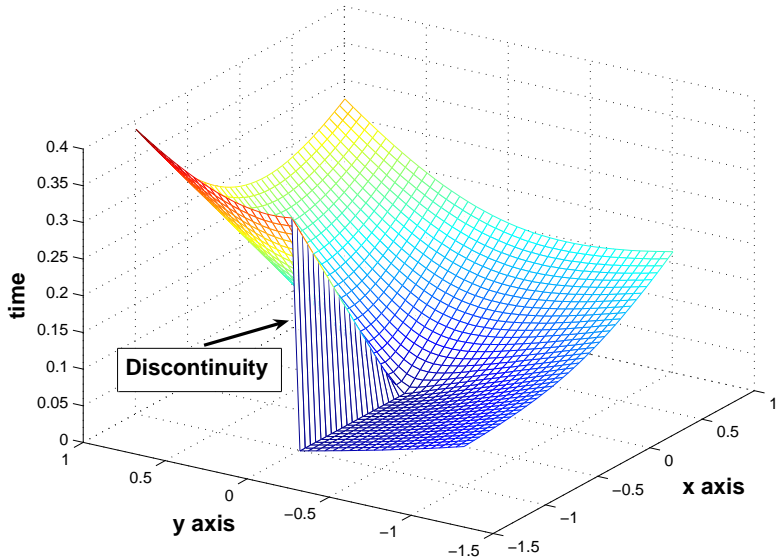}}
\caption{A natural hypersurface for a field with uniform $p_\theta/T_{00}$.  The z-axis is suppressed.}
\label{Figure:corkscrew}
\end{figure}

It will now be demonstrated that such a discontinuity implies a quantization of the angular momentum, at least for field solutions with only non-relativistic frequency components $\omega\approx\omega_0=mc^2/\hbar$.  These solutions are dominated by an oscillation at $\omega_0$, so in the frame $(\bm{x}',t')$ defined by the natural hypersurface $t'=0$, $\phi$ is most usefully specified by the form
\begin{equation}
\label{eq:phiexp}
\phi(\bm{x}',t')=\phi_c(\bm{x}',t') cos(\omega_0 t') + \phi_s(\bm{x}',t') sin(\omega_0 t').
\end{equation}
The warped geometry of a corkscrew hypersurface can generally be avoided by only considering (\ref{eq:phiexp}) valid over some limited 4D volume $0\le t'\le \Delta t$, $|\bm{x}'-\bm{x}_0|\le c\Delta t$.  Here $\Delta t$ will eventually be equated with the temporal discontinuity, and the spatial extent (around some position of interest $\bm{x}_0$) need only be sufficiently large to fully determine the value of the field around ($\bm{x}'=\bm{x}_0$, $t'=\Delta t$) using $\phi$ and its time-derivative on $t'=0$ as specified in (\ref{eq:phiexp}).  The non-relativistic limit can be defined by the maximum wavenumber $k_{max}$ in the Fourier transforms of $\phi_c(t'=0)$ and $\phi_s(t'=0)$; if $k_{max}=\alpha mc/\hbar$, the non-relativistic limit is simply $\alpha \ll 1$.

The restriction $\partial \phi / \partial t'=0$ on the natural boundary condition leads to the condition that $\phi_s/\phi_c<\alpha^2$ everywhere on $t'=0$.  Furthermore, the KGE strongly restricts the evolution of $\phi_c$ and $\phi_s$ towards the point $(\bm{x_0},\Delta t)$, guaranteeing that $\phi_s(\bm{x_0},t') \ll \phi_c(\bm{x_0},t')$ continues to hold throughout the 4-volume of interest, so long as $\alpha\omega_0\Delta t\ll 1$.

Now, consider that the spacetime point $(\bm{x_0},\Delta t)$ lies on the next cycle of the discontinuous hypersurface, such that $\Delta t$ is the period of the ``corkscrew'' at the spatial location $\bm{x}_0$.  The non-relativistic limit guarantees that this next surface will only be slightly tilted with respect to the original $t'$ axis, and in order to enforce $\partial \phi / \partial \eta=0$ on this next surface, the form of (\ref{eq:phiexp}) and the condition $\phi_s(\bm{x_0},t') \ll \phi_c(\bm{x_0},t')$ demands that $\Delta t$ be approximately an integer multiple of $\pi/\omega_0$.  In descriptive language, if the surfaces $t'=0$ and $t'\approx\Delta t$ are both on amplitude ``crests'' or ``troughs'' of a wave solution with a dominant frequency at $\omega_0$, then there is a quantization of $\Delta t$ such that there are roughly a half-integer number of wave periods between the two surfaces.

For the type of discontinuity shown in Figure 1, no global solution is possible if $\alpha\ll 1$.  Such a feat would require that the duration across the discontinuity is always very close to an integer multiple of $\pi/\omega_0$, at \textit{every} spatial location.  This clearly cannot hold, because the temporal gap across the discontinuity varies continuously with $r$.  In the case of an NCBC-measurement on a classical field, therefore, the above conjecture restricts the BC to a particular class of hypersurfaces with a constant-duration discontinuity $\Delta t = n \pi/\omega_0$, where $n$ is an integer.  

Independent of this quantization condition, $\Delta t$ can be calculated in terms of a line integral that lies in the hypersurface, winding from $(\bm{x}_0,0)$ to $(\bm{x}_0,\Delta t)$.  Defining a ``local group velocity'' $\bm{v}$ via the stress energy tensor $v_i\equiv cT_{0i}/T_{00}$ gives the slope and gradient of the ``natural'' hypersurface, because in the frame of such a surface, $\bm{v'}=0$.  So any such line integral on $\bm{s}_0$ advances time by an amount
\begin{equation}
\label{eq:Dt}
\Delta t = \frac{1}{c^2} \oint \bm{v} \cdot \bm{dl},
\end{equation}
where the line integral is a closed loop in 3D, not 4D.  Setting $\Delta t$ equal to $n \pi/\omega_0$ yields a quantization rule
\begin{equation}
\label{eq:BSrule}
m \oint \bm{v} \cdot \bm{dl} \approx  \frac{n}{2} h.
\end{equation}

For the particular case of angular momentum measurements, assuming a cylindrical symmetry around the z-axis, one can write $p_\theta=T_{00}v_\theta/c^2$ and build up the 3D hypersurface $\bm{s_0}$ out of 1D line integrals (each yielding the same quantum number $n$ via (\ref{eq:BSrule})):

\begin{eqnarray}
\label{eq:ptheta}
L_z=\int r p_\theta d^3\bm{s}_0 &=& \int r \frac{T_{00}}{c^2} \left[\int v_\theta r d\theta \right] d^2s_0 \nonumber\\
&\approx& \int \frac{nh}{2mc^2} T_{00} \, r d^2s_0\nonumber\\
&=& \frac{n}{2} \frac{\hbar}{mc^2} \int T_{00} d^3\bm{s}_0\nonumber\\
&\approx& \frac{n}{2} \hbar \, \frac{E_{tot}}{mc^2}
\end{eqnarray}
Here the 3D integration was split up using $d^3{s_0}=rd\theta d^2s_0$, and the full 3D integral was reconstructed using the cylindrical symmetry condition $\int d\theta/(2\pi)=1$.  The final step in (\ref{eq:ptheta}) approximates the previous integral as the total field energy $E_{tot}=\int T_{00} d^3\bm{x}$, which is valid in the limit that only a very small fraction of field energy passes through the discontinuity in $\bm{s}_0$.

\section{Discussion}

This striking result implies an approximate quantization of the classical scalar field angular momentum into units of $\hbar/2$, if such a measurement is imposed as a BC and if the field has a total energy near $mc^2$.  Such an energy would naturally result for any single-particle interpretation of the scalar field, as the derivation of (\ref{eq:ptheta}) already assumed the non-relativistic limit.  Of more general interest is the result (\ref{eq:BSrule}), reminiscent of the Bohr-Sommerfeld quantization condition of the old quantum theory (with an extra factor of $1/2$).  It is difficult to draw too many conclusions here because of fundamental differences between field and particle dynamics, but such a general result seems to open the door to the quantization of energy should external potentials be introduced to this free scalar field.  

The classical scalar field is directly analogous to the quantum field for a neutral spinless particle, for which one might have hoped to find $L_z$ quantized in units of $\hbar$, not $\hbar/2$.  But the only way to physically constrain half-integer values for a scalar field is to somehow have the ``corkscrew`` hypersurface move from ``crests'' ($\phi>0$) to ``troughs''  ($\phi<0$) without violating the  condition $\partial\phi/\partial\eta=0$.  This seems difficult to achieve in practice; if it were impossible, the ``extra'' factor of $1/2$ in both (\ref{eq:BSrule}) and (\ref{eq:ptheta}) would disappear.

These results also shed light on the relationship between measurement and quantization in a classical context.  Note that it is the orientation of the BC/measurement that forces the particular quantization, not anything special about the z-axis.  Also, a simultaneous measurement of a non-zero $L_z$ and $L_x$ is not possible, simply because there is no single ``natural'' hypersurface that can constrain both of these quantities.  Furthermore, if neither side of the hypersurface $\bm{s}$ corresponds to an angular momentum measurement, the quantization requirement disappears, once again allowing a superposition of different angular momenta in $\phi$.  The picture described here is reminiscent of the ``measurement problem'' for quantum systems, but here, of course, there is no interpretation difficulty: everything is purely classical.

Still, despite the classicality, the overall framework might be puzzling, perhaps with an appearance of circular logic.  To recap the situation, one starts with an \textit{a priori} Lagrangian density ${\cal L}$ of an unknown field $\phi$.  One then constrains $\phi$ on $\bm{s}$, the closed hypersurface boundary of some spacetime volume $\Omega$; this is a physically realized BC, imposed via physical measurements/interactions external to $\Omega$.  The resulting field solution $\phi(\Omega)$ is one which both conforms to the BC on $\bm{s}$ and also extremizes the action ($S=\int {\cal L} d\Omega$) under variations of $\phi$ that are consistent with the BC.   The extremization requirement, however, implies that not all BCs (and not all hypersurfaces $\bm{s}$) are physically realizable. 

Fortunately, there is no circular logic here; the BC determines $\phi(\Omega)$, not the other way around; the BC (including the shape $\bm{s}$) is constrained not by any particular solution $\phi(\Omega)$, but rather the global constraint that the action must be extremized \textit{somehow}.  Indeed, the global constraint on $\bm{s}$ depends only on $\partial\phi/\partial\eta |_{\bm{s}}$, not $\phi(\Omega)$, and the former is precisely what the NCBCs constrain in the first place.  

In summary, by taking Hamilton's principle seriously -- even when using constraints for which it was not originally intended -- one necessarily finds that classical fields have certain quantized properties.  For example, when using a scalar field with total energy near $mc^2$, measurements of the angular momentum would only yield results near $n\hbar/2$, where $n$ is an integer.  This quantization is not exact; any given measurement would have a small range of allowed values around the quantized peaks (on the order of $\alpha^2n^2\hbar$).  Although this result is merely meant as an example, it does imply that experimental tests of the above conjecture may certainly be possible.  More definitely, these results indicate (along with other, related developments \cite{Dolce, Wharton, Pirsa}) that the search for realistic interpretations of quantum phenomena, based entirely upon classical field theory, remains a promising avenue of research.

\section*{Acknowledgments}
The central conjecture in this paper resulted from a stimulating discussion with D. Dolce concerning his manuscript \cite{Dolce}.  The author also thanks A. de la Fuente, R. Sutherland and R. Spekkens for insightful comments.  This research was supported in part by the Perimeter Institute for Theoretical Physics.

\end{document}